\documentclass{PoS}

\title{Few-nucleon scattering experiments\\ at intermediate energies}

\ShortTitle{Few-nucleon scattering experiments}

\author{\speaker{Johan Messchendorp}\\
        Kernfysisch Versneller Instituut, University of Groningen, Zernikelaan 25, 9747 AA, Groningen, The Netherlands\\
        E-mail: \email{messchendorp@kvi.nl}}

\abstract{Observables in few-nucleon scattering processes are sensitive probes
to study the two and many-body interactions between nucleons in nuclei. 
In the past decades, several facilities provided a large data base 
to study in detail the few-nucleon interactions below the pion-production 
threshold by exploiting polarized proton and deuteron beams and
large-acceptance detectors. In this contribution, some recent results 
are discussed and interpreted by rigorous Faddeev calculations which are based upon 
modern potentials. Furthermore, the paper presents 
preliminary results from a pioneering and exclusive study of the four-nucleon 
scattering process at intermediate energies which was recently conducted 
at the KVI.}

\FullConference{6th International Workshop on Chiral Dynamics, CD09\\
		July 6-10, 2009\\
		Bern, Switzerland}

\begin{document}

\section{Introduction}

Understanding the exact nature of the nuclear force is one of the
long-standing questions in nuclear physics. In 1935, Yukawa
successfully described the pair-wise nucleon-nucleon (NN) interaction
as an exchange of a boson~\cite{Yukawa}. Current NN models are mainly
based on Yukawa's idea and provide an excellent description of the
high-quality data base of proton-proton and neutron-proton
scattering~\cite{stoks94} and of the properties of the
deuteron. However, for the simplest three-nucleon system, triton,
three-body calculations employing NN forces clearly underestimate the
experimental binding energies~\cite{wiringa95}, demonstrating
that NN forces are not sufficient to describe the three-nucleon system
accurately. Some of the discrepancies between experimental data and
calculations solely based on the NN interaction can be resolved by
introducing an additional three-nucleon force (3NF). Most of the
current models for the 3NF are based on a refined version of
Fujita-Miyazawa's 3NF model~{\cite{fuji}}, in which a 2$\pi$-exchange
mechanism is incorporated by an intermediate $\Delta$ excitation of
one of the nucleons~{\cite{deltuva03II,coon01}}. More
recently, NN and three-nucleon potentials have become available 
which are derived from the basic symmetry properties of 
the fundamental theory of Quantum Chromodynamics (QCD)~\cite{chiptnn1,chiptnn2}. 
These so-called chiral-perturbation ($\chi$PT) driven models construct systematically
a potential from a low-energy expansion of the most general
Lagrangian with only the Goldstone bosons, e.g. pions, as exchange
particles. The validity of the $\chi$PT-driven models for the intermediate energies
discussed in this paper remains, however, questionable and depends strongly on
the convergence of results at higher terms in the momentum expansion.


\section{Nucleon-deuteron elastic scattering}

\begin{figure}[ht]
\centering 
\includegraphics[width=\textwidth]{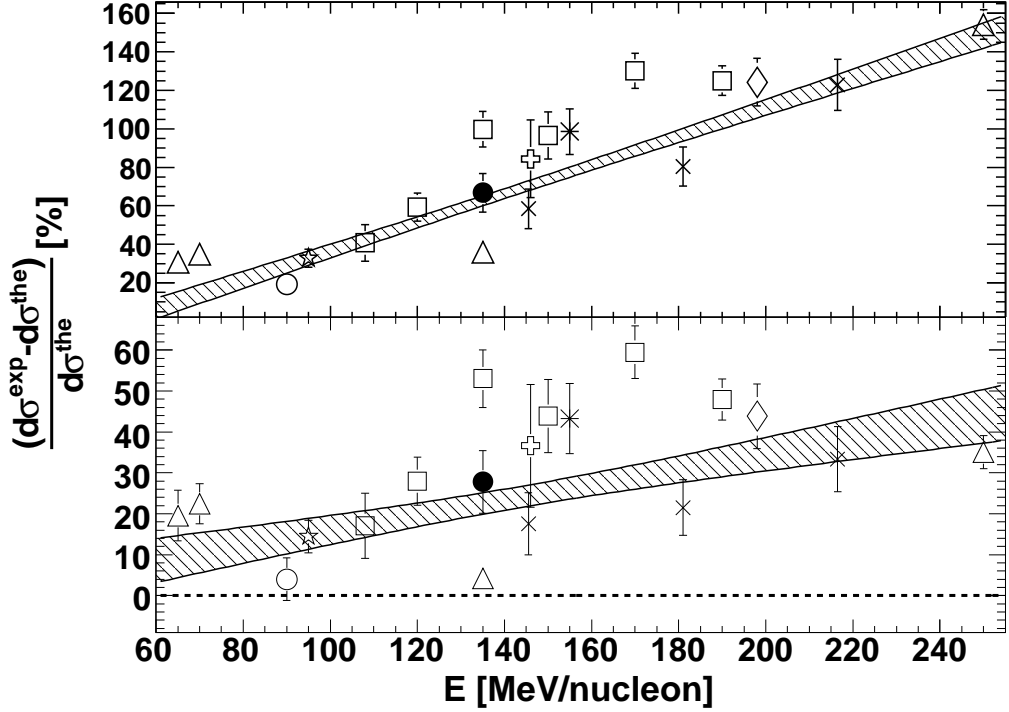}
\caption{The relative difference between the calculations by the Hannover-Lisbon 
theory group and the measured cross sections for the elastic $p+d$ reaction as a
function of beam energy for $\theta_{\rm c.m.}=140 ^\circ$. 
The top panel shows the differences with a calculation based
on the CD-Bonn potential and the Coulomb interaction, whereas for the 
bottom panel an additional $\Delta$ isobar has been taken into account.
Open squares are data from Ref.~{\cite{kars03}}, open triangles are data from
Refs.~\cite{sakai00,kimiko02,shimi,hatan}, open circle is from~{\cite{hamid}},
open star is from~{\cite{mermod}}, crosses are from~{\cite{Igo}}, star
is from~{\cite{kurodo}}, open cross is from~{\cite{postma}}, diamond is from
~{\cite{ald}} and the filled circle is from~{\cite{ram08}}.  The shaded
band represents the result of a line fit through
the data excluding the results obtained at KVI, RIKEN and RCNP. 
The width of the band corresponds to a 2$\sigma$ error of the fit.}
\label{syscheck} 
\end{figure}

In the last decade, high-precision data at intermediate
energies in elastic \it {Nd} \rm and \it {dN} \rm
scattering~{\cite{kars01,kars03,kars05,bieber,sakai00,kimiko05,
kimiko02,postma,hamid,kurodo,mermod,Igo,ald,Hos,Ela07,shimi,hatan,IUCF}}
for a large energy range together with rigorous Faddeev
calculations~{\cite{gloeckle}} for the three-nucleon system have
proven to be a sensitive tool to study the 3NP. In particular, a large
sensitivity to 3NF effects exists in the minimum of the differential
cross section~{\cite{witala98,nemoto}}. 
The results of a systematic study of the energy dependence of 
all available cross sections in elastic proton-deuteron scattering 
with respect to state-of-the-art calculations by the Hannover-Lisbon 
theory group are depicted in Fig.~\ref{syscheck}. The top panel shows 
the relative difference between the model predictions 
excluding the $\Delta$-isobar contribution and data taken at a fixed 
center-of-mass angle of $\theta_{\rm c.m.}$=140$^\circ$. The data points 
were extracted from a polynomial fit through each angular distribution. 
The error bars correspond to a quadratic sum of the statistical and systematic
uncertainties of each measurement. Note that the discrepancies, reflecting
the 3NF effects, increase drastically with incident energy and reach 
values of more than 100\% at energies equal or larger than 200~MeV.
The bottom panel in Fig.~\ref{syscheck} shows a similar comparison
between data and model predictions including the $\Delta$-isobar as
mediator of the 3NF effects. Clearly, a large part of the discrepancies
is resolved. However, a smaller but significant deficiency remains which
increases with energy to values of about 30\% at an energy of 200~MeV.

\section{Nucleon-deuteron break-up}

Complementary to the elastic scattering experiments, three-nucleon
studies have been performed exploiting the nucleon-deuteron break-up
reaction. The phase space of the break-up channel is much richer than that of
the elastic scattering. The final state of the break-up reaction is
described by five kinematical variables, as compared to just one for the
elastic scattering case. Therefore, studies of the break-up reaction
offer a way of much more detailed investigations of the nuclear
forces, in particular of the role of 3NF effects. Predictions show that 
large 3NF effects can be expected at specific kinematical regions in the
break-up reaction. Results of the cross sections and tensor analyzing
powers have already been published for a deuteron-beam energy of 130~MeV 
on a liquid-hydrogen target~{\cite{kistryn06,ola06,ela07}}. These
experiments were the first ones of its type which demonstrated the 
feasibility of a high-precision measurement of the break-up observables and they
confirmed that sizable influences of 3NF and Coulomb effects are
visible in the break-up cross sections at this energy. In the last
years, more data at several beam energies and other observables have
been collected to provide an extensive data base at intermediate
energies.

\begin{figure}[ht]
\centering
\includegraphics[angle=0,width=0.6\textwidth]{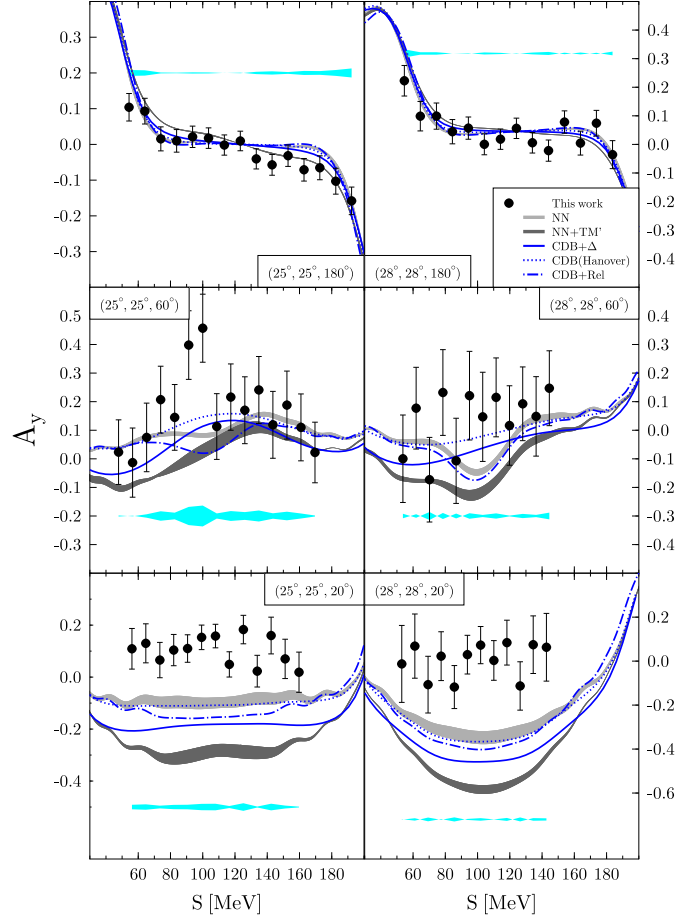}
\caption{A comparison between the results of the analyzing power
 measurements for a few selected break-up configurations with various
 theoretical predictions. The light gray bands are composed of various
 modern two-nucleon (NN) force calculations, namely CD-Bonn, NijmI,
 NijmII, and AV18. The dark gray bands correspond to results of the
 calculations with the same NN forces including the TM' (3N)
 potential.  The lines represent the predictions of calculations by
 the Hannover-Lisbon group based on the CD-Bonn potential (dotted) and
 CD-Bonn potential extended with a virtual $\Delta$ excitation (solid blue). 
 The blue dash-dotted lines are derived from calculations by
 the Bochum-Cracow collaboration based on the CD-Bonn potential
 including relativistic effects~\cite{skib06}. The errors are statistical and the
 cyan band in each panel represents the systematic uncertainties
 (2$\sigma$).}
\label{crossall}
\end{figure}

Recent and interesting results have been obtained at KVI using a 4$\pi$
detection system BINA, which provides a unique tool to study a large
part of the phase space of the break-up reaction. 
Figure~\ref{crossall} presents some preliminary results of the vector
analyzing powers in proton-deuteron break-up for an incident proton
beam of 190~MeV and for two symmetric kinematical configurations 
$(\theta_{1},\theta_{2})$=($25^{\circ},25^{\circ}$) and 
($28^{\circ},28^{\circ}$) for three different values of $\phi_{12}$. 
Here, the angles $\theta_1$ and $\theta_2$ refer to the polar angles
of the two final-state protons and $\phi_{12}$ to the relative azimuthal
angle between these protons. The parameter $S$ is directly related to the
energies of the two final-state protons and is a measure of their energy
correlation. The data are compared with
calculations based on different models for the interaction dynamics as
described in detail in the caption of the figure. For these configurations and
observable, the effects of relativity and the Coulomb force are
predicted to be small with respect to the effect of three-nucleon
forces. At $\phi_{12}$=$180^\circ$, the value of $A_y$ is predicted
to be completely determined by two-nucleon force effects with only a
very small effect of 3NFs, which is supported by the experimental
data. Note, however, that the effect of 3NFs increases with decreasing
of the relative azimuthal angle $\phi_{12}$, corresponding to a
decrease in the relative energy between the two final-state protons.
The observed discrepancies could point to a
deficiency in the spin-isospin structure of the description of the
many-nucleon forces in the present-day state-of-the-art calculations
as discussed in Ref.~\cite{mar09}.

\section{Exclusive deuteron-deuteron break-up}

The 3NF effects are in general small in the three-nucleon system. A
complementary approach is to look into systems for which the 3NF
effects are significantly enhanced in magnitude. For this, it was
proposed to study the four-nucleon system. The experimental data base
in the four-nucleon system is presently poor in comparison with the
three-nucleon system.  Most of the available data were taken at very
low energies, in particular below the three-body break-up threshold of
2.2~MeV. Also, theoretical developments are evolving rapidly at low
energies~\cite{RamazaniA_Ref23,RamazaniA_Ref24,RamazaniA_Ref25,RamazaniA_Ref26},
but lag behind at higher energies. The experimental data base at
intermediate energies is very
limited~{\cite{RamazaniA_Ref27,RamazaniA_Ref28,RamazaniA_Ref29}}.
This situation calls for extensive four-nucleon studies at
intermediate energies. 

\begin{figure}[ht]
\centering
\includegraphics[width=0.8\columnwidth,angle=0]{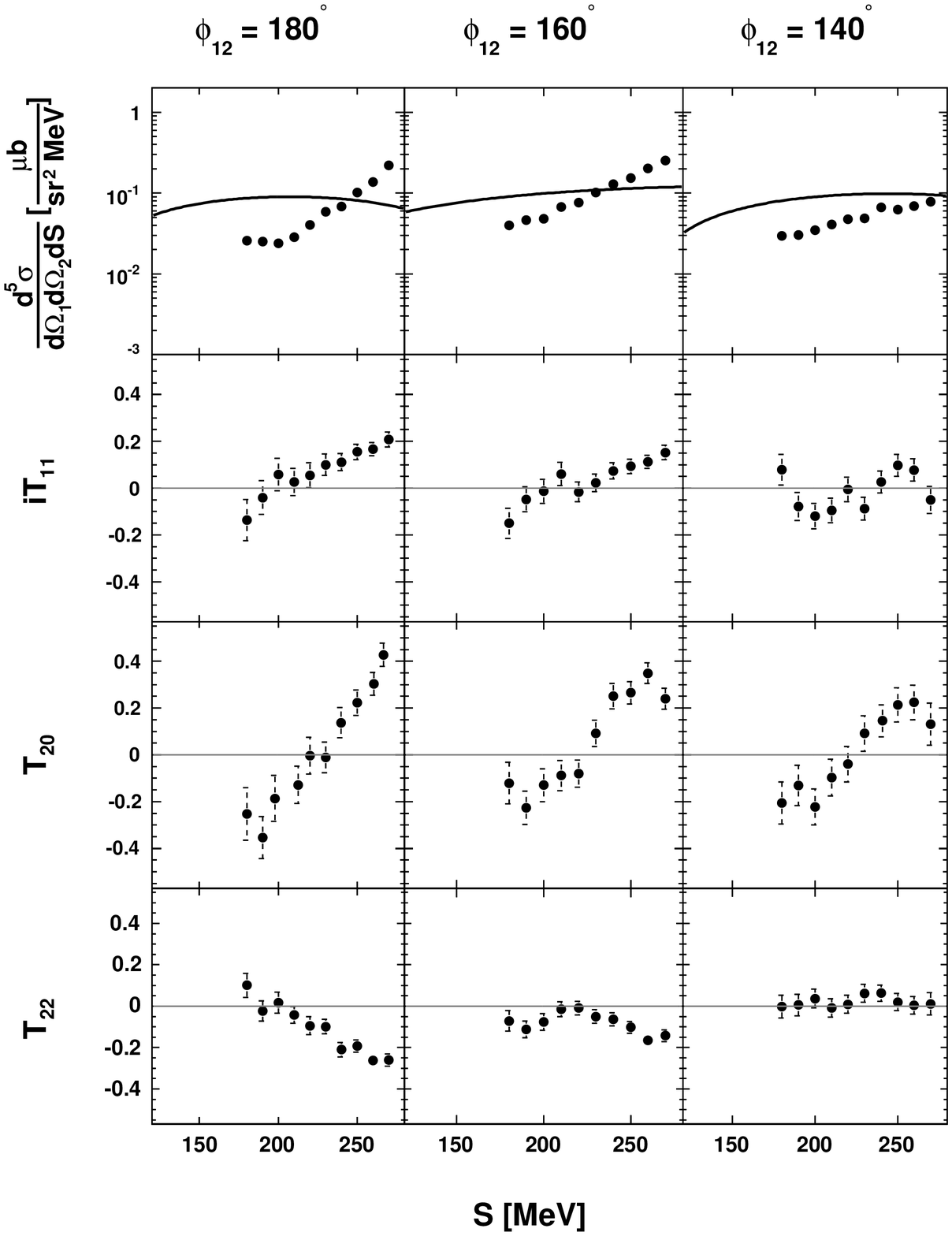}
\caption{The cross sections, vector-, and tensor-analyzing
 powers at $(\theta_{d}, \theta_{p}) = (15^{\circ}, 15^{\circ})$ as a
 function of $S$ for different azimuthal opening angles. The solid
 curves in the top panels correspond to phase-space distributions.
 They have arbitrary normalization with respect to the data.  The gray
 lines in other panels show the zero level of the analyzing powers.
 Only statistical uncertainties are indicated.}
\label{dd1}
 \end{figure}

Recently, comprehensive measurements of 
cross sections and spin observables in various $d+d$ scattering processes 
at 65~MeV/nucleon, namely the elastic and three-body break-up channels, 
were performed at KVI using the BINA detector. With the corresponding results, 
the four-nucleon scattering data base at intermediate energies is significantly enriched. 
Figure~\ref{dd1} depicts some of the preliminary results of the deuteron-deuteron
three-body break-up reaction, $d+d\rightarrow d+p+n$, which were obtained via the 
unambiguous detection of a proton in coincident with a deuteron in the final state. 
For the first time, a systematic and exclusive study of the three-body
break-up reaction in deuteron-deuteron scattering at intermediate energies
was shown to be feasible and provided precision results in the
four-nucleon sector as well.  

\section{Conclusions}

In the past decades, our understanding of the nuclear forces has
drastically improved. These developments can be attributed to the
enormous progress made in theory and in experiment. In particular,
in the three-nucleon sector, the theoretical descriptions are 
ab-initio, based on high quality potentials, and (partly) able
to include effects like Coulomb and relativity. Also, the
experimental techniques have significantly improved in the course
of time and have provided a huge data base with high-precision
data and covering a huge part of the phase space. The
four-nucleon data base at intermediate energies is growing
significantly, thereby providing potentially new insights 
and a testing ground for our present understanding of the 
many-body force effects.

In spite of the progress made in experimental and theoretical techniques to
study the many-nucleon system, there are still various open questions
which urgently need to be addressed. A large part of these questions
point to our present understanding of 3NF effects. This paper discusses
some results of few-nucleon scattering experiments taken at intermediate energies. 
Although, the overall comparison between data 
and theory improve significantly by taking into account 3NF effects, 
there are still various channels, phase spaces, and observables which 
show huge discrepancies. Therefore, the existing data base for few-nucleon
scattering observables provide an ideal basis to develop a better understanding of
three-nucleon force effects in few-nucleon interactions.

\section*{Acknowledgments}

The author acknowledges Mohammad Eslami-Kalantari, Hossein Mardanpour, and 
Ahmad Ramazani. The results presented here are part of their PhD theses. Furthermore, 
the author thanks Nasser Kalantar for the valuable discussions and his input. 
This work is part of the research program of the Stichting voor Fundamenteel Onderzoek 
der Materie (FOM) with financial support from the Nederlandse Organisatie 
voor Wetenschappelijk Onderzoek (NWO). The present work 
has been performed with financial support from the University of Groningen and the
GSI, Helmholtzzentrum f\"ur Schwerionenforschung GmbH, Darmstadt.

\end{document}